\documentstyle{elsart}

\begin{document}
\begin{frontmatter}
\title {Chaos and Statistical Mechanics in the Hamiltonian 
Mean Field model}

\author{Vito Latora\thanksref{vito}}
\address{Center for Theoretical 
Physics, Laboratory for Nuclear Sciences
and Department of Physics, Massachusetts Institute 
of Technology, Cambridge, Massachusetts 02139, USA} 

\author{Andrea Rapisarda\thanksref{andrea}}
\address{Istituto Nazionale 
di Fisica Nucleare, Sezione di Catania
and Dipartimento di Fisica, Universit\'a di Catania,
Corso Italia 57, I-95129 Catania, Italy } 

\author{Stefano Ruffo\thanksref{stefano}}
\address{Dipartimento di Energetica ``S. Stecco", 
Universit\'a di Firenze and INFN, Via S. Marta, 
3 I-50139, Firenze, Italy} 
 
\thanks[vito]{E-mail: latora@ctp.mit.edu}
\thanks[andrea]{E-mail: andrea.rapisarda@ct.infn.it}
\thanks[stefano]{and INFN Firenze, E-mail:ruffo@ing.unifi.it}

\begin{abstract} 
We study the dynamical and statistical behavior
of the Hamiltonian Mean Field (HMF) model in order to investigate 
the relation between microscopic chaos and phase transitions. HMF is 
a simple toy model of 
$N$ fully-coupled rotators which shows a second order phase transition.
The canonical thermodynamical solution is briefly recalled and its 
predictions are tested numerically at finite $N$. The Vlasov
stationary solution is shown to give the same consistency equation
of the canonical solution and its predictions for rotator angle and momenta
distribution functions agree very well with numerical
simulations. 
A link is established between the behavior of the maximal Lyapunov
exponent and that of thermodynamical fluctuations, expressed by
kinetic energy fluctuations or specific heat.
The extensivity of chaos in the $N \to \infty$ limit
is tested through the scaling properties of Lyapunov spectra and 
of the Kolmogorov-Sinai entropy.
Chaotic dynamics 
provides the mixing property in phase space 
necessary for obtaining equilibration; however, the relaxation time to 
equilibrium grows with $N$, at least near the critical point.
Our results constitute an 
interesting bridge between Hamiltonian chaos in 
many degrees of freedom systems and 
equilibrium thermodynamics. 
\end{abstract}

\begin{keyword}
Hamiltonian dynamics, equilibrium statistical
mechanics, Lyapunov exponents, relaxation to equilibrium 
\centerline{\bf Accepted for publication in Physica D}
\end{keyword}

\end{frontmatter}

\section{Introduction}
Many-particle systems  can show collective behavior when the 
average kinetic energy is small
enough. This collective macroscopic behavior can coexist
with chaos at the microscopic level. Such a behavior is particularly
evident for systems that have a phase transition, for which
a nonvanishing order parameter measures the degree of macroscopic
organization, while at the microscopic level chaotic motion
is a source of disorder.
The latter  can induce non trivial time dependence in the macroscopic
quantities, and it would be desirable to relate the time behavior
of such quantities and their fluctuations to the chaotic properties 
of microscopic motion, measured through the Lyapunov spectrum.
A naive idea is that an increase of chaos as the energy
(temperature) is increased should be accompanied with a growth of
fluctuations of some macroscopic quantity. These should be
maximal at the critical point and then drop again at high
energy.
In this paper we study a model of $N$ fully-coupled Hamiltonian
rotators which realizes such a behavior, it has been called
Hamiltonian Mean Field (HMF) model~\cite{antoni,latora}.
It can also be considered as a system of interacting particles
moving on a circle. 
This system has a second order phase transition and in the
ordered phase the rotators are clustered; the high temperature
phase is a gaseous one, with the particles uniformly distributed
on the circle.
It has been shown in ref.~\cite{latora} that the maximal 
Lyapunov exponent grows up to the critical energy density $U_c$ and 
then drop to zero in the whole high temperature phase in the 
$N \to \infty$ limit.
Correspondingly one observes a growth of kinetic energy 
fluctuations up to the critical point and then a phase of vanishing 
fluctuations.
Finite $N$ effects complicate this simple picture. In the
high temperature phase the maximal Lyapunov exponent vanishes 
quite slowly (with $N^{-1/3}$) and  finite size effects
influence the first region below the critical point. In this region
the system displays metastability: starting far from equilibrium,
this is reached in a time $\tau_r$ which grows with $N$.
On the contrary, the extremely low energy phase is characterized
by a weak $N$ dependence, with the maximal Lyapunov exponent $\lambda_1$
which behaves as $\lambda_1 \sim \sqrt{U}$.

Although the model is extremely simplified, it shares many features
with more complex models, for which the relation between chaotic motion
at the microscopic level and collective macroscopic properties
has been studied. Let us mention studies in solid state physics
and lattice field theory~\cite{solid,nayak,dellago,yama,lapo}.
However, it has been actually in nuclear physics~\cite{ata,cmd}, 
where there is presently a lively debate on multifragmentation 
phase transition~\cite{ata,cmd,gsi,eos,bondgro,bond,perco,cmd1,mastinu}, 
that the interest in the connection between chaos and phase transitions 
has been revived. In this case in fact,
an energy/temperature relation quite close to the HMF model has been 
observed~\cite{gsi} and critical exponents have been measured 
experimentally~\cite{eos}. 
Statistical thermodynamical models~\cite{bondgro} and 
percolation approaches ~\cite{perco}
have been proved to give a good description of the experimental 
data, though  the dynamics is missing. 
On the other hand classical molecular dynamics
models \cite{cmd,bond,cmd1} seem to contain all the main ingredients, but 
have the disadvantage that a detailed 
 understanding of the dynamics  can be too complicated.
In this respect, the HMF model can be very useful in clarifying 
some general dynamical features which could be eventually compared 
with real experimental data.
In fact, when studying nuclear multifragmentation, one deals with 
excited clusters of  100-200 particles interacting via long-range 
(nuclear and Coulomb) forces.
Quantum effects are relevant only at very low
energy. In fact in the nuclear case, at very low energy, $T$ is not 
linear in $U$, but grows as $\sqrt{U}$ 
because nucleons are fermions~\cite{gsi,bondgro}.  
However, a classical picture should be quite realistic in the critical
region where the excitation energy is substantial ~\cite{mastinu}.

In this paper we present new numerical data concerning both statistical
quantities, like specific heat and distribution functions, and chaotic
probes, like Lyapunov spectra and Kolmogorov-Sinai entropy. Moreover,
we add to the theoretical analysis of the model a thorough treatment
of differences in the fluctuating quantities between the canonical and
microcanonical ensembles. We also 
investigate in detail the relaxation to equilibrium and compare 
numerical results with a complete self-consistent Vlasov calculation
of distribution functions. Finally a  comparison of numerically obtained
maximal Lyapunov exponents with theoretical formulas is attempted.

The paper is organized as follows. In Sec. 2 we  briefly discuss 
the details of the HMF model. The equilibrium statistical mechanics and 
the continuum Vlasov solution are described in Sects. 3 and 4 
respectively. In Sec. 5 we discuss the relaxation to equilibrium
and in Sec. 6 we present the numerical calculations of 
the Lyapunov spectra and Kolmogorov-Sinai entropy as a function of 
the energy and $N$. Analytical estimates are discussed in Sec. 7 
and conclusions are drawn in Sec. 8.

\section{The HMF model}
The Hamiltonian we consider is the following
\begin{equation}
H(\{ \theta_i \},\{  p_i \}) = K + V~, 
\end{equation}
\noindent
where 
\begin{equation}
K=\sum_{i=1}^N {1\over2} p_i^2,  ~ ~ ~
V = {\epsilon\over 2N} \sum_{i,j=1}^N [ {1-cos(\theta_i-\theta_j)} ]
\end{equation}
\noindent
are the kinetic and potential energies. The model describes the motion of N
rotators characterized by the angle $\theta_i \in [0, 2 \pi)$: 
each rotator interacts with all the others.
One can define a spin vector associated with each rotator 
$
{\bf m}_i = ( cos(\theta_i), sin(\theta_i) ) ~.
$
\noindent
The Hamiltonian then describes N classical spins similarly to the XY model, 
and a ferromagnetic or an antiferromagnetic behavior according 
to the positive or negative sign of $\epsilon$ ~\cite{antoni}. 
In the following we will consider only the ferromagnetic (attractive) 
case and moreover we put $\epsilon=1$ without loss of generality.
After defining
$
{\bf M} =   {1\over N} \sum_{i=1}^N {\bf m}_i = (M_x, M_y)~,
$
the potential energy $V$ can be rewritten as
\begin{equation}
V =   {N\over 2} (1 -  (M_x^2 + M_y^2))  ~= 
      {N\over 2} (1 -  M^2 )            ~.
\end{equation}
The equations of motion then read
\begin{equation}
\dot{\theta_i}  =   p_i ~~~~,~~~~
\dot{p_i}  = -sin(\theta_i) M_x   + cos(\theta_i) M_y ~~~,~~~ i=1,...,N~.
\label{eqmoto} 
\end{equation}
The evolution equations of the tangent vector are
\begin{equation}
\label{lin1}
\dot{\delta \theta_i}  =   \delta p_i ~~~~,~~~~ 
\dot{\delta p_i}=  
-\sum_j {{\partial^2V}\over {\partial \theta_i \partial \theta_j} } 
\delta \theta_j~~,  
~~~~ i=1,...,N.
\end{equation}
\noindent
where the diagonal and off-diagonal terms of the Jacobian
$J_{ij}= - \partial^2 V /\partial \theta_i \partial \theta_j$ are 
\begin{equation}
\label{der1}
J_{ii}=-cos(\theta_i) M_x - sin(\theta_i) M_y  + {1\over{N}}
\end{equation}
\begin{equation}
\label{der2}
J_{ij}
~= {1\over{N}}cos(\theta_i-\theta_j)~~~,    ~~~i \neq j
~~.
\end{equation}
\noindent
Expression (\ref{der1}) can also be written for convenience as: 
\begin{equation} 
\label{diag}
J_{ii}=-M cos(\theta_i-\phi) + \frac{1}{N} 
\end{equation}
\noindent
where $\phi$ is the phase of {\bf M }.

The HMF model was initially proposed in Ref~\cite{antoni}. A time
discrete version of it was previously introduced in Ref.~\cite{kaneko} 
in the form of globally coupled Chirikov standard
maps. The Lyapunov instability of the HMF has been first studied
numerically in Ref.~\cite{yama}, and analytically in
Ref.~\cite{firpo} using a Riemaniann geometry approach~\cite{lapo}. 
The connection between Lyapunov instability
and thermodynamical properties has been established in Ref.~\cite{latora}.
In the original proposal of the HMF~\cite{antoni}, the relation of 
the model with self-gravitating systems (in the attractive case)
and charged sheets models (in the repulsive case) was studied; this 
has been also taken over for the gravitational case in 
Ref.~\cite{ina} and in the context of plasma models~\cite{elskens1,elskens2}. 
It has also been shown that the HMF is the peculiar representative
of a larger class of models, whose thermodynamics can be solved
exactly~\cite{elskens1}. Moreover, anomalous diffusion properties
of a generalization of the model, with two angles for each rotator,
have been recently studied~\cite{tor}.

\section{Canonical and microcanonical results}
It is interesting to look at the predictions of statistical
mechanics. We restrict here to the $N \to \infty$
limit, where microcanonical and canonical ensemble results
coincide for averaged quantities (apart from exceptions near 
first order phase transitions, see Refs.~\cite{thir}). 
The free energy of our model can be easily obtained in the canonical 
ensemble. This was done in Ref.~\cite{antoni} and the result reads
\begin{equation}
-\beta F = \frac 1 2 \log \biggl(\frac {2 \pi} {\beta} \biggr)
-\frac {\beta}{2} + \max_y \biggr(- \frac {y^2} 
{2 \beta}
+\log \Bigr(2 \pi I_0 (y) \Bigl) \biggl)~,
\label{enerlib}
\end{equation}
where $\beta=1/k_B T$ (the Boltzmann constant $k_B$ is set to 1) 
and $I_i$ is the 
modified Bessel function of $i$-th order. The auxiliary 
variable $y$ is introduced to 
decouple the particles by the usual Hubbard-Stratonovich trick
for Gaussian integrals, and the search of the maximum in 
Eq.~(\ref{enerlib}) (which is a consequence of the continuum limit 
$N \to \infty$ solved with the saddle point method)
leads to the consistency equation
\begin{equation}
\frac y {\beta}-\frac{I_1}{I_0}(y)=0~.
\label{cons}
\end{equation}
The magnetization $M=I_1/I_0$ is obtained by solving the consistency
equation, which is done numerically. It vanishes for 
$\beta < \beta_c = 2$, the inverse critical temperature.
At low temperatures it approaches $1$ (the limit of $I_1/I_0$).

The energy-temperature relation (sometimes called the {\em caloric
curve} ) is
\begin{equation}
U = \frac{E}{N} = \frac{\partial (\beta F)}{\partial \beta}=
\frac{1}{2 \beta} + \frac{1}{2} ( 1 - M^2)~,
\end{equation}
thus $U_c= E_c/N= 3/4$.
 
Close to the critical point $\beta \to \beta_c^+$, magnetization
and energy behave as
\begin{eqnarray}
M & \approx & \frac{4}{\beta} \sqrt{\frac{1}{2} -
\frac{1}{\beta}} \\
U & \approx & \frac{1}{2 \beta} \bigl[ 1 - \frac{8(\beta
-2)}{\beta^2} \bigr] + \frac{1}{2}~.
\end{eqnarray}
Hence, $M$ vanishes with the $1/2$ classical mean field exponent and the
specific heat $C_V = \partial U/\partial T$ is finite at the critical
point $C_V(\beta_c)=5/2$ and constant ($C_V=1/2$) 
in the high temperature phase.
Thus for what concerns the critical behavior,  
$C_V \sim (T_c- T)^{-\alpha}$ with $\alpha=0$.

These theoretical results are compared with those of numerical
simulations in Fig.~1, where care is taken to use almost equilibrated
initial data in order to reduce the relaxation time 
(as discussed in Sects. 4 and 5). 
We have integrated Eqs.~(\ref{eqmoto}) using fourth order 
symplectic algorithms~\cite{yo} with a time step $\Delta t \sim 0.2$, 
adjusted to maintain the error in energy conservation below  
${\Delta E\over E} =10^{-5}$.
The agreement is good over the whole energy range and finite N effects,
although present, are weak.

In the inset of Fig.~1b we plot the results obtained starting from
non equilibrated initial data (the ``water bag" initial condition
discussed in Sect. 5) and although the integration time was quite
long ($O(10^5)$), a sharp disagreement is observed just below the
critical point.
A region of negative specific heat is present and a
continuation of the high temperature phase (linear $T$ vs. $U$ relation) 
into the low temperature one (metastability). It is very intriguing 
that these out-of-equilibrium quasi-stationary states (QSS) show a caloric 
curve very similar to the one found for first order phase transitions
~\cite{tor,thir,laba,gross}.
In that case, however, the corresponding states are equilibrated and
do not die asymptotically, as it is shown in Sec.~5 for our QSS.
In our case, at equilibrium, we do not have phase-coexistence, all rotators 
belong to a single cluster (see Sec.~4); phase-coexistence can
arise only as a finite $N$ non-equilibrium effect.
The existence of long-living non-equilibrium states has been 
noticed already in Refs.~\cite{latora,yama} and a connection to critical
slowing down has also been proposed. More recently
it has also been found numerically in other one-dimensional models~\cite{ppp} and
in self-gravitating systems~\cite{miller}, but in this case it is
not associated to the closeness of a phase transition. 
The coexistence of different states in the continuum limit near the 
critical point is a purely microcanonical effect. It arises because of the 
inversion of the $t \to \infty$ limit with the $N \to \infty$ one.

Concerning fluctuations, it is well known that 
the predictions of microcanonical ensemble
differ from those of the canonical~\cite{verlet}. For instance,
while in the canonical ensemble kinetic energy fluctuations are given by
\begin{equation}
\Sigma_{can}=\frac{\sigma_{can} (K)}{\sqrt{N}} = 
\sqrt{\frac{(<K^2>_{can} -<K>_{can}^2)}{N}}=
\frac{1}{\sqrt{2} \beta}~,
\end{equation}
their expression in the microcanonical ensemble is~\cite{firpo}
\begin{equation}
\label{fluct}
\Sigma_\mu=\frac{\sigma_\mu(K)}{\sqrt{N}}  =  \frac{T_\mu}{\sqrt{2}} 
\sqrt{ 1 -  \frac{1}{ 1 - 2 M ({dM}/{dT_\mu}) } }
\end{equation}
\noindent
where $T_\mu= 2 <K>_\mu/N$ (for a rigorous definition of microcanonical
temperature see~\cite{rugh}). The latter is compared with numerical simulations 
in Fig.~5b. The specific heat is~\cite{verlet}
\begin{equation}
\label{verlet}
C_V^\mu = \frac{1}{2} \left( 1 - 2 \left( \frac{\Sigma_\mu}{T_\mu}
\right)^2 \right)^{-1}~,
\end{equation}
and also compares quite well with numerical simulations (see Fig.~5c);
although finite $N$ effects are obviously stronger for fluctuations 
than for the averages.

\section{The Vlasov solution}

The Vlasov equation for our system reads
\begin{equation}
\frac{\partial f}{\partial t} + p \frac{\partial f}{\partial \theta}
- \frac{\partial V}{\partial \theta} \frac{\partial f}{\partial p}=0~,
\label{Vlasov}
\end{equation}
where $f(\theta,p,t)$ is the normalized distribution function and
the potential $V$ satisfies the equation
\begin{equation}
\frac{\partial^2 V}{\partial \theta^2} =
\int_0^{2 \pi} \int_{-\infty}^{\infty} \cos (\theta - \theta')
f(\theta',p',t) d \theta' d p'~.
\label{potential}
\end{equation}
A simple hypothesis that can be made on the distribution function $f$ is
that it factorizes as
\begin{equation}
f=f_0(p)g(\theta,t)~.
\label{factor}
\end{equation}


Although simple, this is a very strong hypothesis and its validity can
at present be justified only by the correctness of the results which
are derived thereby (see the following). In the high temperature 
phase one expects a uniform distribution in $\theta$ and then $f$ 
should depend only on $p$, this fact
is then consistent with the factorization hypothesis. Lowering the 
temperature below the transition point, a modulation in $\theta$
appears. This is a manifestation of the Jeans instability (see
Ref.~\cite{ina}). If the modulation is not too strong,
the factorization hypothesis is again reasonable. However, in the
low temperature/energy phase this hypothesis has no clear {\it a priori} 
justification. 


By further requiring that $f_0$ is Gaussian
\begin{equation}
f_0= \frac{1}{\sqrt{2 \pi T}} \exp - \frac{p^2}{2T}~,
\label{Gaussian}
\end{equation}
one gets the equation
\begin{equation}
\frac{\partial g}{\partial t} + p \frac{\partial g}{\partial \theta}
- \frac{p}{T}\left( M_y \cos \theta - M_x \sin \theta \right)
g(\theta)=0
\end{equation}
where $M_x= \int \cos \theta g d\theta$, $M_y= \int \sin \theta 
g d\theta$. 

Restricting to the stationary solution, 
$\partial g/\partial t=0$ one can easily solve for $g(\theta)$
\begin{equation}
g=g_0 \exp \left[ \frac{1}{T} \left( M_y \sin \theta +M_x \cos
\theta \right) \right] = g_0 \exp \left[ \frac{M}{T}
\cos (\theta - \phi) \right]~,
\label{g}
\end{equation} 
where $g_0=1/(2 \pi I_0 (M/T))$ as imposed by normalization
and $\phi$ is the phase of ${\bf M}$. It must be observed that
$g$ is expressed in terms of $(M_x,M_y)$ or $(M,\phi)$, which 
are themselves functions of $g$; so one must solve the problem 
self-consistently, after writing the equations for the two components
of the magnetization
\begin{eqnarray}
M_x &=& \cos \phi \frac{I_1(M/T)}{I_0(M/T)} \label{cons1}\\
M_y &=& \sin \phi \frac{I_1(M/T)}{I_0(M/T)}~.
\label{cons2}
\end{eqnarray}
These equations coincide exactly with the consistency equations
(\ref{cons}) of the solution in the canonical ensemble. 
Once $(M_x,M_y)$ are determined by solving (numerically) 
Eqs.~(\ref{cons1},\ref{cons2}), they
can be substituted back into Eq.(\ref{g}), thus obtaining the
stationary distribution function. Due to the global phase
translation invariance of the model, for any $M$ the choice 
of the phase $\phi$ is arbitrary, as reflected in the solutions
of Eqs.~(\ref{cons1},\ref{cons2}). 
With respect to the results in Ref.~\cite{ina} we have fully
determined the distribution function by imposing the self-consistency
conditions through  Eqs.~(\ref{cons2}).

In the high temperature phase ${\bf M}=0$ thus $g=1/(2 \pi)$ is
uniform. 

At very low temperature one can use the asymptotic
development of $I_0$
\begin{equation}
I_0 (z) = \frac{\exp (z)}{\sqrt{2 \pi z}} \left[ 1 + \frac{1}{8z} + 
\dots \right]
\end{equation}
to get the Gaussian
\begin{equation}
g \sim \frac{1}{\sqrt{2 \pi \sigma^2}} \exp \left( - \frac{\theta^2}
{2 \sigma^2} \right)~,
\end{equation}
where $\sigma^2= T/M$ is the variance, which
vanishes with $T$, giving a Dirac-$\delta$ at zero temperature.

A comparison of formulas (\ref{Gaussian},\ref{g}) with numerical 
data is shown in Fig.~2; the theoretical curve fits the data very
accurately with no free parameter.
Both in the low energy region (Fig.~2a,b) and at higher energy, where
the cluster drifts (Fig.~2c,d), the agreement is very good.
The theory does not determine the value of $\phi$, which remains
arbitrary; so we have adjusted this value to the center of the 
cluster, which is moving in time quite irregularly.

\section{Slow relaxation to equilibrium}

Around the critical energy, relaxation to equilibrium depends in a 
very sensitive way on the adopted initial condition~\cite{mick}.
When starting with ``water bag'' initial conditions, i.e.
a flat distribution function of finite width centered around zero 
for $f_0(p)$, and putting all rotators at $q_i=0$ 
($g(\theta)= \delta(0)$), we reveal
the presence of quasi-stationary (long living) non-equilibrium states
(QSS) (see the inset of Fig.~1b).
In Fig.~3 the evolution to equilibrium of the QSS state is shown
by the time evolution of the reduced distribution function $f_0(p,t)$.
It is only at $t=5\cdot10^5$ that a good reproduction of the 
Gaussian distribution and a convergence to the predicted 
equilibrium temperature $T_\mu=0.4757$ for $U=0.69$ is obtained for $N=1000$.
Such a slow relaxation is observed in the region just below the
critical point (see Fig.~1b) and also around $U \sim 1$.
In order to study the $N$-dependence of the relaxation time $\tau_r$ we have
roughly quantified the distance from the equilibrium state by measuring
\begin{equation}
\Delta S = |S(t) - S^{eq}|~,
\end{equation}
where $S= -\int f_0(p,t) \ln f_0(p,t) dp$ is the Boltzmann entropy
of the momentum reduced distribution function, and $S^{eq}$
its equilibrium value when the distribution is a Gaussian at the
given equilibrium temperature $T_\mu$.
The results, shown in Fig.~4, clearly indicate an increase of the
relaxation time with $N$ (Fig.~4(a)), which can be approximately
fitted with a linear law $\tau_r \sim N$.
The convergence to the equilibrium value of $S$ is not exact at
finite $N$ and the error decreases as $N^{-1/2}$; this is shown
in Fig.~4(a) by the convergence to a decreasing value (horizontal
lines) as $N$ is increased from 1000 to 10000.
A similar law was found in Ref.~\cite{antoni} by studying the time
needed to reach equipartition of rotators velocity in the large
$U$ region and in Ref.~\cite{elskens2} analysing the time needed
to absorb holes in the momentum distribution (so called Dupree 
structures in plasma physics) for the antiferromagnetic HMF.
It then seems that various indicators agree in suggesting a
diverging time scale with $N$. However, the time scale
also depends on $U$, and what we have here found is that
it is much greater in the region near the critical point.
This could be a manifestation of critical slowing down.

\section{Lyapunov spectra and Kolmogorov-Sinai Entropy}

We have computed the Lyapunov spectrum $\lambda_i \ ,  i=1,\dots,N$ 
by the standard method of Ref.~\cite{ben}. The average 
number of time steps in order to get a good convergence was of 
the order $10^6$. We discuss in the following 
numerical results for system sizes in between N=10 and N=20000. 

In Fig.~5(a) we plot $\lambda_1$ as a function of $U$ for various $N$
values. As the system is integrable both in the limit of very small 
and very large energies (reducing in the former case to weakly
coupled harmonic oscillators and in the latter to free rotators), the
maximal Lyapunov exponent must vanish in these two limits. 

In the region of weak chaos, for  $U < 0.25$, the curve has
a weak $N$-dependence. Then $\lambda_1$ changes abruptly and   
a region of stronger chaos begins.
In Ref.~\cite{antoni} it was observed that in between $U=0.2$ and $U=0.3$ 
a different dynamical regime sets in and particles start to evaporate 
from the main cluster. A similar regime was found in Ref.~\cite{cmd} and 
this behavior is  also similar to the one found in Ref.~\cite{nayak} 
at a solid-liquid transition. 
In this region of strong chaoticity we observe a pronounced 
peak already for $N=100$~\cite{yama}, which persists and becomes broader
for $N=20000$.

The standard deviation  $\Sigma_\mu$ of the kinetic energy fluctuations 
is plotted in Fig.~5(b) and it compares quite
well with the theoretical prediction (\ref{fluct}), although finite $N$ 
effects are larger than for averaged quantities.
In Fig.~5(c) we report the microcanonical specific heat obtained
with formula~(\ref{verlet}), compared with numerical simulations 
at increasing values of $N$.
Both these quantities display a similar behavior to the maximal
Lyapunov exponent: they increase up to the critical point and
then drop to zero. It is therefore quite natural to associate
the growth of the Lyapunov exponent to the growth of fluctuations,
as expressed both by the kinetic energy fluctuations and the
specific heat. In this respect a similar connection was proposed in 
Ref.~\cite{aldoprep}.
Unfortunately, a theoretical formula for the Lyapunov exponent
as $N \to \infty$ does not yet exist (see anyway next section); 
however, the data in Fig.~5(a) already show what we should expect 
for the convergence of $\lambda_1$ as $N$ increases.
For $U \ge U_c$, $\lambda_1 \to 0$ as $N \to \infty$, thus revealing
the presence of a whole region of integrability in this limit; rotators
decouple, reducing the system to a ``gas" of free rotators (which is 
consistent also with the vanishing of kinetic energy fluctuations).
Fig.~6(a) shows that the convergence to zero can by fitted as
$\lambda_1 \sim N^{-1/3}$. This scaling law can be derived theoretically
using a random matrix approximation~\cite{latora} (see also next
section), which is also shown in the same figure to approximate
quite well the numerical results at large enough energy.
(A $N^{-1/3}$ convergence to the asymptotic value of the Lyapunov
exponent is also observed in systems of hard spheres \cite{hard}).
The question is still open whether $\lambda_1$ will show a 
discontinuity at $U_c$ in the $N \to \infty$ limit, as the kinetic
energy fluctuations and specific heat do.
Strong finite size effects are present in the region $U \in [0.2,0.75]$.
This is the region where the cluster drifts (see~\cite{antoni}), while
particles evaporate from and condensate on it. The Lyapunov exponent is
here significantly larger than at smaller energies. For $U < 0.2$,
we observe a fast convergence to the $N \to \infty$ limit, and the scaling
law $\lambda_1 \sim \sqrt{U}$ is numerically obtained (see Fig.~6(b)).
A heuristic justification of this scaling was proposed in 
Ref.~\cite{latora}, and a new derivation is presented in Sec.~ 7.1.

The extensivity of the Lyapunov spectrum was first proposed by 
Ruelle~\cite{ruelle} and numerically tested in Refs.~\cite{spec}
(see~\cite{ruffo} for a review). It amounts to check that plotting
$\lambda_i$ vs. $i/N$ and letting $N$ go to infinity, while keeping
fixed physically intensive parameters (in our case energy density 
or temperature), one obtains a convergence to an asymptotic curve,
so called distribution of Lyapunov exponent.
This is verified for our model in Fig.~7(a) at various energy 
densities. The asymptotic Lyapunov distribution is more 
quickly reached, as $N$ increases, at smaller energies 
(confirming the fast $N$ convergence observed above for the maximal 
Lyapunov exponent at small energy). 
The data at $U=0.7$ show a much weaker convergence to the asymptotic
spectrum, as also confirmed by looking at the Kolmogorov-Sinai entropy
density at the same value of $U$ (upper points in the inset of Fig.~8).
Heavier numerical simulations are needed to assess the convergence
of the spectrum in this energy region.
The spectra have a curious exponential shape in the first part 
(see the inset of Fig.~7(a)), which was already noticed in Ref.~\cite{kaneko}.
No significative change in the shape of the spectra at $N=50$ and
$N=100$ is observed when going from below to above the phase 
transition point (see Fig.~7(b) for the $N=100$ case).
However, we cannot exclude that differences in the spectra could arise
at larger values of $N$, which are so far inaccessible to numerical
experiments.

The Kolmogorov-Sinai (K-S) entropy density (which, due to Pesin's
formula, is in our case the sum of the positive Lyapunov
exponents divided by $N$), $S_{KS}/N$, is plotted in Fig.~8 against
$U$. It shows again a peak at $U_c$, a fast convergence to a
limiting value as $N$ increases in the small energy limit and a
slow convergence to zero (as expected) for $U \geq U_c$.
The scaling laws are here roughly: $S_{KS}/N \sim U^{3/4}$ at small $U$,
see Fig.~9(a), and $S_{KS}/N \sim N^{-1/5}$ for $U>U_c$, see Fig.~9(b)
(but a more refined numerical analysis is needed to confirm these 
results).

Therefore, the analysis of the behavior of the Lyapunov spectrum
and of the K-S entropy confirms the picture already emerging from
the study of the maximal Lyapunov exponent, showing an increase
of microscopic chaos near the phase transition point.
It is somewhat intriguing that it is precisely when chaos is stronger
that one reveals a slowing down of the relaxation to equilibrium (see
Sec. 5); although not unexpected, because it is quite symplistic 
to associate the notion of chaos to that of efficient diffusion 
of orbits in phase space. When starting from a ``water bag" initial
condition one reaches the almost frozen QSS state, and also for
this state the transient Lyapunov exponent has been checked to be
positive, although slightly different than the asymptotic value in
the equilibrium state. It could be that the phase space near the
transition point has a rich structure with many coexisting chaotic 
QSS states.

\section{Analytical estimates of the maximal Lyapunov exponent}

In this section we discuss some theoretical approaches which
allow to justify the scaling laws of the maximal Lyapunov exponent
observed in numerical experiments, and also to obtain some
order of magnitude estimates.

\subsection{Using the Vlasov solution}

One can try to use the stationary Vlasov distribution 
function (\ref{Gaussian},\ref{g}) to derive some properties of the 
tangent space. For instance, averaging the Jacobian $J_{ij}$ 
(\ref{der1},\ref{der2}) over this
distribution, one obtains a constant matrix whose diagonal
and off-diagonal elements are
$-M^2+ 1/N$ and $M^2/N$, respectively (we are
here also neglecting correlations among the particles). It is then
quite easy, being now the Jacobian a constant matrix,
to compute the maximal Lyapunov exponent. The
result is
\begin{equation}
\lambda_1 = \sqrt{\frac{1 - M^2}{N}},
\label{Lyap-Vlasov}
\end{equation}
with Lyapunov eigenvector $(\{ \delta \theta_i \}, \{ \delta  p_i \})= 
({\bf a},{\bf b})$,
the vectors ${\bf a}$ and ${\bf b}$ being constant. This formula has the 
correct dependence on $M$, in fact at small energies the virial relation
$<K> \sim <V>$ holds and then $U=E/N \sim 1-M^2$, giving 
$\lambda_1 \sim \sqrt{U}$, which is what is observed numerically.
However, formula (\ref{Lyap-Vlasov}) predicts that the Lyapunov exponent
vanishes with $N^{-1/2}$, which is not observed in numerical experiments.
We have found numerically that, even if we allow for the true temporal
fluctuations of $M^2$, we obtain for the Lyapunov exponent the same value
of formula (\ref{Lyap-Vlasov}) and the same Lyapunov eigenvector.
On the contrary, if we look at the Lyapunov eigenvector given by the 
true dynamics, we observe that it is far from being constant, only a few
components being significantly different from zero. We have checked
that the number of nonvanishing components of the Lyapunov eigenvector
remains constant as $N$ increases (this has also been recently
found also for a generalization of the HMF studied in Ref.~\cite{tor}). 
Thus, the typical size of each component of the vector remains constant 
as $N$ grows (remember that eigenvectors 
are normalized), while for constant eigenvectors this size decreases 
as $N^{-1/2}$; this is a naive argument that can explain the extra 
$N^{-1/2}$ present in formula (\ref{Lyap-Vlasov}) with respect to 
what is observed for the true Lyapunov, i.e. $\lambda_1 \sim \sqrt{U}$.
We have recently developed a perturbation theory scheme, which shows 
that for sparse eigenvectors $\lambda_1$ does not vanish with $N$
at small energies and, with some additional hypotheses, one obtains the
$\sqrt{U}$ law~\cite{new}.

\subsection{A numerical test of a recently proposed formula}

We compare in this subsection a recently proposed formula for
the maximal Lyapunov exponent~\cite{lapo,firpo} with numerical 
results, giving also an intuitive interpretation 
of it for our model.
In Ref.~\cite{lapo} a general method is formulated
to describe Hamiltonian chaos using the differential geometric 
structure underlying the dynamics. 
They consider the Hamiltonian many-body dynamics 
as a geodesic flow on a Riemannian manifold; then the  
chaotic motion reflects the instabilities of the 
geodesic flow, which depend on the curvature properties 
of the manifold.
They obtain the following formula for $\lambda_1$
\begin{equation}
\label{casettiformula}
\lambda_1= {\Lambda\over 2} - {{2 \Omega_0 } \over {3\Lambda} } ~~;~~~
\Lambda = \left( 2 \sigma^2_{\Omega} \tau 
+ \sqrt{ \left( {{4 \Omega_0}\over{3}} \right)^3 
+ {\left( 2 \sigma^2_{\Omega}  \tau \right)}^2} \right)^{1\over 3}~,
\end{equation}
where $\Omega_0$ and $\sigma^2_{\Omega}$ are the average Ricci
curvature and the variance of its fluctuations respectively.
For the HMF, these two quantities turn out to be given by very simple 
expressions in terms of the averages and fluctuations 
introduced in Sec.~3 (see also ref.~\cite{firpo})
\begin{eqnarray}
\label{omegazerohmf}
\Omega_0          &=& <M^2> - \frac{1}{N}  
                   =  T_\mu + (1-2U) - \frac{1}{N}
\nonumber\\
\sigma^2_{\Omega} &=&   N  {\sigma^2_{M^2}} 
                   =    4 \Sigma_\mu^2~,
\end{eqnarray}
where $\sigma^2_{M^2}$ is the variance of the fluctuations
of $M^2$. All this is the consequence of the fact that the
Ricci curvature,$-1/N \sum_i J_{ii}$ with $J_{ii}$ given by 
formula (\ref{der1}), is simply given in our model
by $M^2 -1/N$.
The Ricci curvature is assumed in Ref.~\cite{lapo} to be 
a $\delta$-correlated stochastic process in time, and
the correlation time $\tau$ in formula (\ref{casettiformula})
is estimated as
\begin{equation}
\label{tau}
{\tau}^{-1}= 2( {\tau_1}^{-1} +  {\tau_2}^{-1})~; 
~~~
{\tau_1}= \frac {\pi } { 2 \sqrt{\Omega_0+\sigma_{\Omega}}   }~; 
~~
{\tau_2}= \frac {\sqrt{\Omega_0}} {\sigma_{\Omega}}
\end{equation}
Some comments about the derivation and the use of formula 
(\ref{casettiformula}) are very important.
It is derived in the ``diagonal" approximation, which corresponds 
to neglect the effect of off-diagonal terms (\ref{der2}) 
in Eqs.~(\ref{lin1}). This approximation has been checked numerically
not to be valid at small energies, leading to a value of $\lambda_1$
which is orders of magnitude less than the true one (see also Fig.~10).
The assumption of a $\delta$-correlated stochastic process makes 
easier the calculation of $\lambda_1$, but reduces the range of 
applicability of formula (\ref{casettiformula}). 
The estimate for $\tau$ is a rather delicate problem, where
some arbitrariness can enter the theory ~\cite{mehra}.

In the high energy phase, $M$ fluctuates above zero and scales
with $N^{-1/2}$ at large $N$, then we have 
$<M^2> \sim \sigma_{M^2} \sim N^{-1}$ and moreover $\tau \approx
\tau_2/2$ since  $\tau_{2}= M / (\sigma_{M^2}\sqrt{N})=O(1)$
while $\tau_1 \sim N^{1/4}$. Hence one finds ~\cite{firpo}
\begin{equation}
\lambda_1 \sim \Lambda \sim {(4 N \sigma^2_{M^2}\tau)}^{1/3}
\sim N^{-1/3}~.
\end{equation}

In the low energy phase $\sigma_{M^2} \ll {M^2}$ 
and  $\lambda_1$ reduces to
\begin{equation}
\label{formulaprl}
\lambda_1 \sim  \frac{4 \Sigma_\mu^2 \tau}{M^2}
\end{equation}
which gives a relation betwen $\lambda_1$ and the fluctuations 
of kinetic energy. This result supports the link between 
chaos and kinetic energy fluctuations already claimed in Ref.\cite{latora}
although the formula derived heuristically there was different, but
in better agreement with the scaling law for small $U$. 

In Fig.~10 we compare the numerically computed values of  
$\lambda_1$ at increasing values of $N$ with those obtained 
from formula (\ref{casettiformula}), using both the 
 finite $N$ numerical values for $<M^2>$ and $\Sigma_\mu$
and the thermodynamic limit values in the microcanonical
ensemble. We have also checked that 
eqs. (30) are well reproduced by numerical simulations.
Formula (\ref{casettiformula}) reproduces the behavior of $\lambda_1$ 
vs. $U$ within a factor of two over a large range of energies. 
In particular it predicts correctly a maximum of $\lambda_1$ 
around $U_c$ and the $N^{-1/3}$-law in the high energy phase.

However, in the limit $U \to 0$ this formula gives $\lambda_1 \sim U^2$,
as can be easily derived from formula (\ref{formulaprl}), 
which is sharply in contrast with the behaviour $\lambda_1 \sim \sqrt{U}$ 
observed in numerical simulations (see Fig.~6(b)). 
We think that, in this limit, the stochastic approximation for
the average Ricci curvature breaks down, because this quantity
is in our case nothing but $M^2$, and this is an almost
regularly oscillating quantity as $U \to 0$, due to the   
collective excitations of the cluster. However, 
it is possible to use $\tau$ as a fitting parameter to 
reproduce the numerical data.
In fact some preliminary
numerical investigations have shown that in this region the correlation
time given by eq. (31) strongly underestimates the realistic one.

\section{CONCLUSIONS}

We have investigated the dynamical and statistical behavior 
of  a system with long-range forces showing a second order phase transition.
Both the maximal Lyapunov exponent $\lambda_1$ and the Kolmogorov-Sinai 
entropy density $S_{KS}/N$ are peaked at  the phase transition point, 
where kinetic energy fluctuations and specific heat are maximal. There is
actually a small shift to lower energies due to finite size effects.
The latter    
are present also in the Lyapunov spectra and 
in the Kolmogorov-Sinai entropy.
Above the phase transition point, both $\lambda_1$ and $S_{KS}$ 
vanish as $N \to \infty$.
We think that this toy model contains some important ingredients to 
understand the behavior of macroscopic order parameters when dynamical 
chaos is present at the microscopic level. Most
of our findings are probably common to other Hamiltonian systems 
showing second order phase transitions.
In particular our results 
could be very important in order to understand the 
relaxation to the equilibrium solution and 
the success of statistical approaches in describing 
the  nuclear multifragmentation phase transition. 

\begin{ack}
We thank M.C. Firpo for communicating us her results before publication
and P. Holdsworth for interesting suggestions.
We thank A. Torcini for many useful discussions and a careful
reading of the text.
A.R. thanks the Centre for Theoretical Physics of MIT for the kind 
hospitality and M. Robnik for stimulating discussions during his 
visits  at CAMTP in Maribor, Slovenia.
V.L. and S.R. thank INFN for financial support.
S.R. thanks CIC, Cuernavaca, Mexico for financial 
support. This work is also part of the European contract 
No. ERBCHRXCT940460 on ``Stability and universality in classical 
mechanics".

{\it In the 60's, Boris Chirikov was also an explorer of the  (no man's 
land at that time) relation between chaotic motion and statistical behavior
in classical systems with many degrees of freedom. We hope that
he will be interested by this work.}
\end{ack}

\newpage


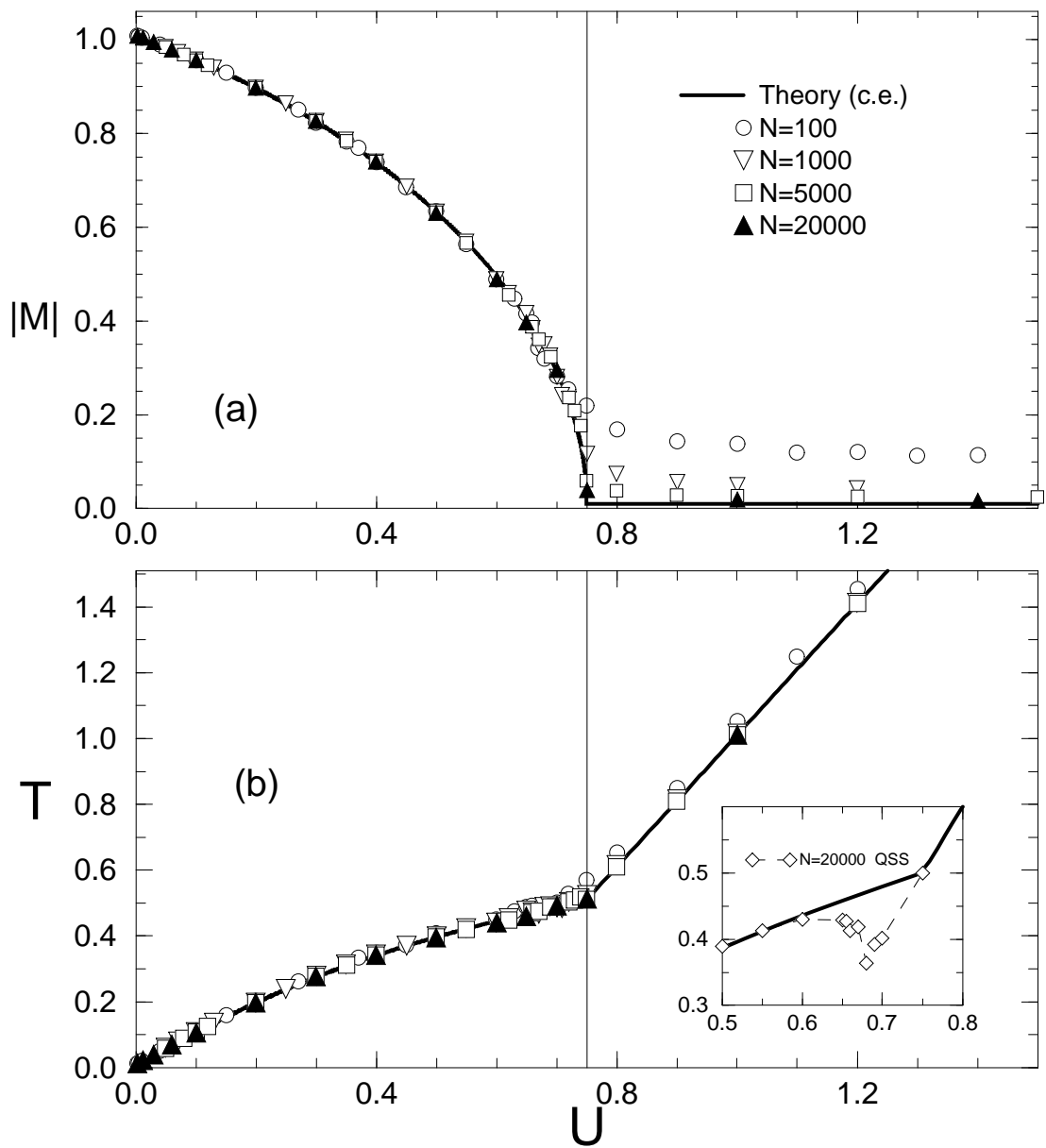
\begin{figure}
\caption{Theoretical predictions in the canonical ensemble  
(full curve) for the magnetization $M$ vs. $U$, panel (a), and  
for the caloric curve $T$ vs. $U$, panel (b), in comparison with numerical 
simulations (microcanonical ensemble) for N=100,1000, 5000, 20000. The 
vertical line indicates the critical energy $U_c=3/4$.
We plot also the microcanonical results for the Quasi-Stationary States 
(QSS) in the case N=20000 (losanges) in the inset of panel (b).}
\end{figure}

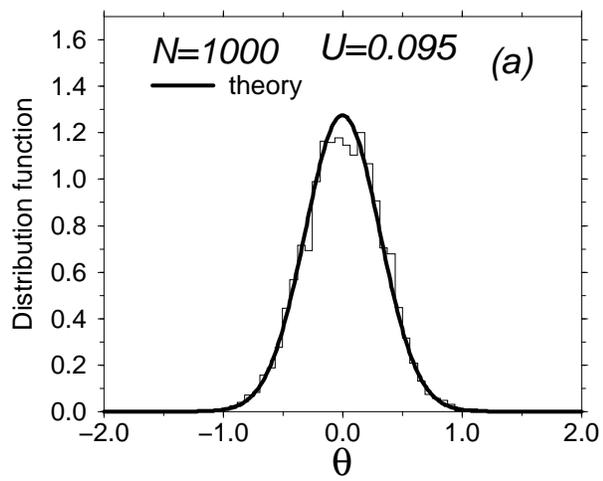
\begin{figure}
\caption{
Equilibrium reduced distribution functions 
in angles $\theta$ and momenta $p$ for $U=0.095$ ((a) and (b))
and $U=0.36$ ((c) and (d)). The numerical simulations for $N=1000$ 
(histogram) are compared with the theoretical distributions
(\ref{Gaussian},\ref{g})
}
\end{figure}

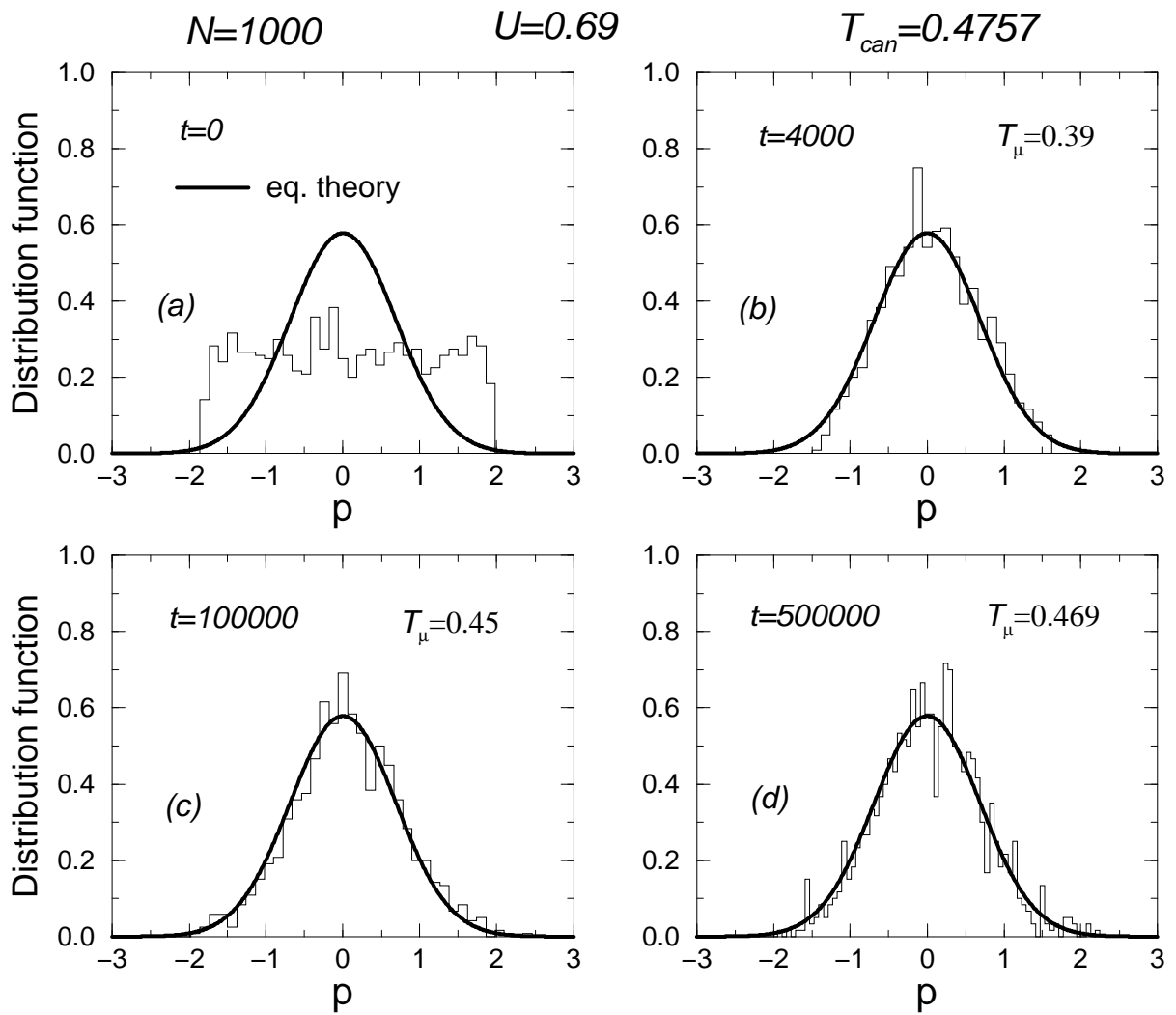
\begin{figure}
\caption{Reduced distribution function in $p$ for
$U=0.69$ and $N=1000$ at increasing times.
The initial distribution is a water bag (histogram). 
The thick full line is the equilibrium distribution.}  
\end{figure}

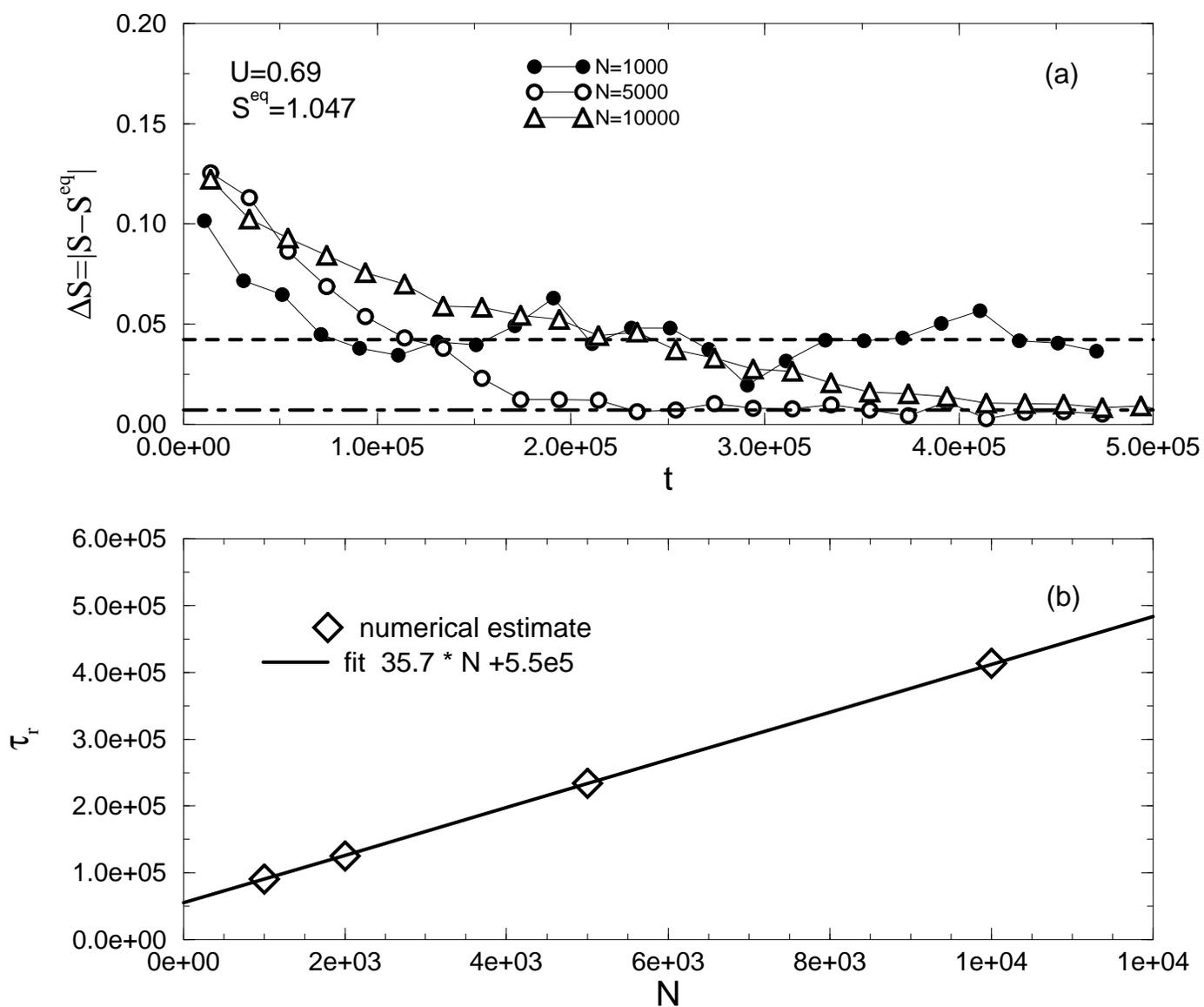
\begin{figure}
\caption{(a) Relaxation to equilibrium of the Boltzmann entropy
of the reduced distribution in $p$ for N=1000 and N=5000; (b)
Linear increase with $N$ of the time scale $\tau_r$ for the relaxation to
equilibrium}  
\end{figure}

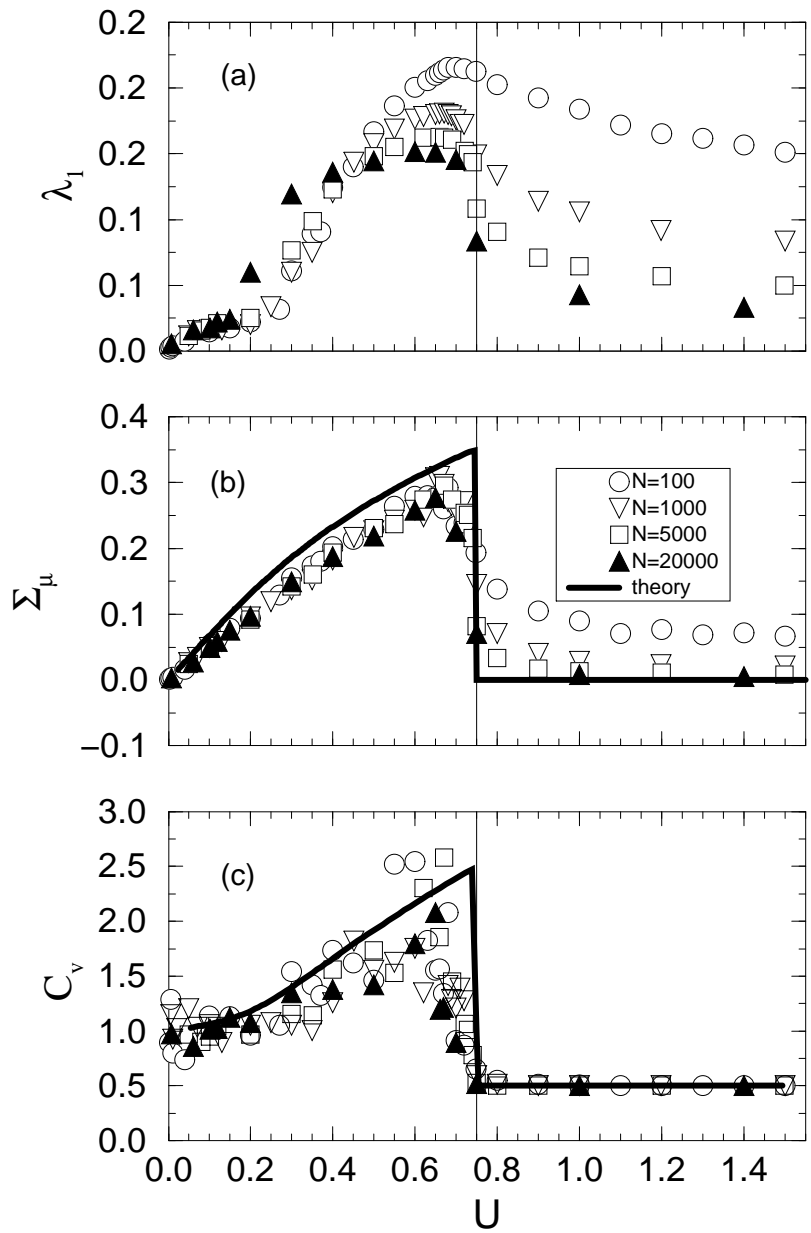
\begin{figure}
\caption{(a)Numerical data for the largest Lyapunov exponent as 
a function of $U$ for various system sizes: N=100,1000, 5000 
and 20000. (b) Kinetic energy fluctuations $\Sigma_\mu$ vs. $U$.
(c) Specific heat obtained from formula (\ref{verlet}).
The vertical line indicates the critical energy $U_c=3/4$. 
The full lines in panels (b) and (c) are the microcanonical 
predictions.}
\end{figure}

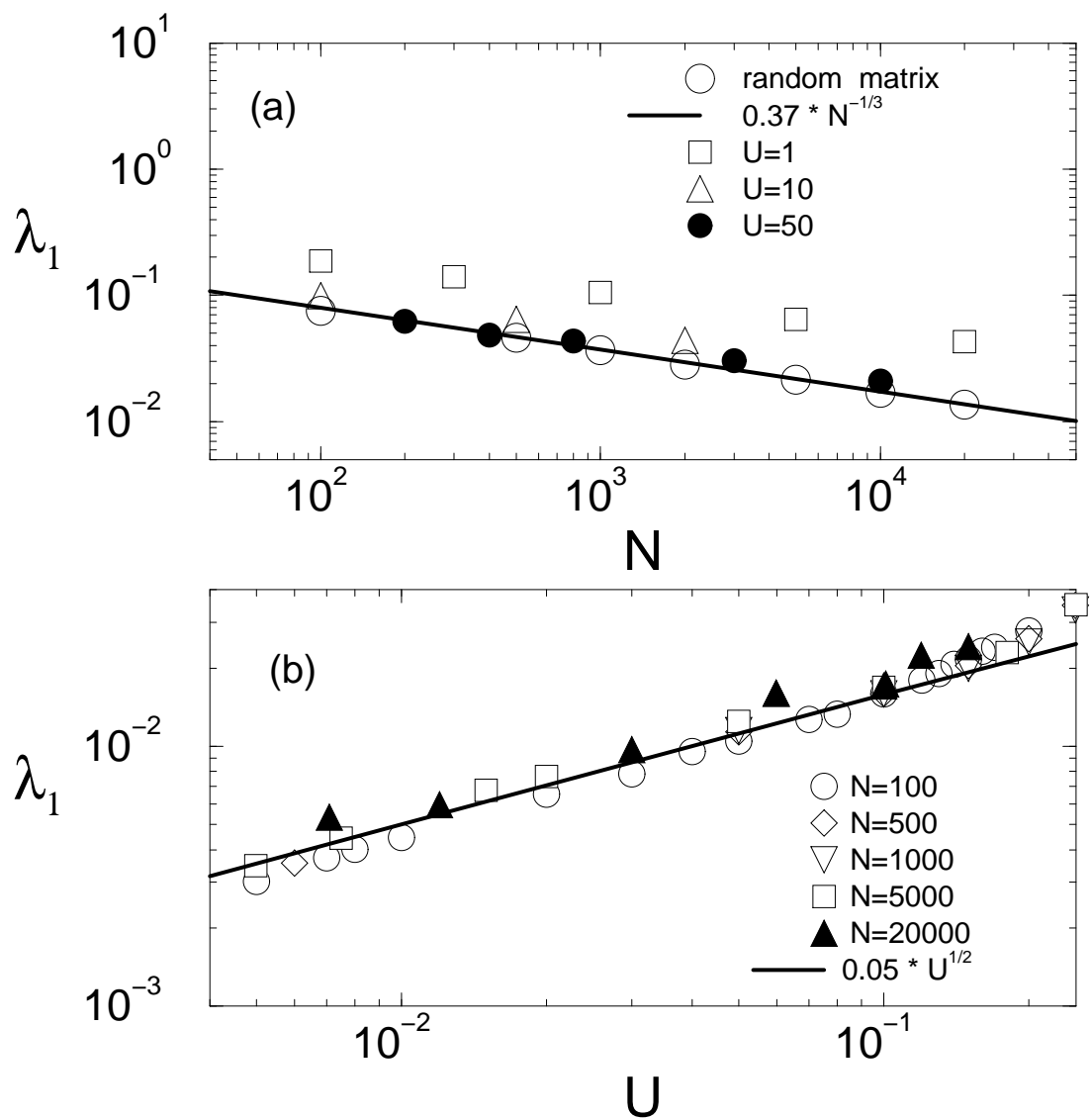
\begin{figure}
\caption{(a)Convergence to zero of the largest Lyapunov exponent vs. 
$N$ in the high energy region. The fitted curve is the power law
$N^{-1/3}$, the circles are the result of a random matrix simulation. 
(b) $U^{1/2}$ growth of the maximal Lyapunov exponent vs. $U$
at small $U$, a rather weak $N$ dependence is observed.}
\end{figure}

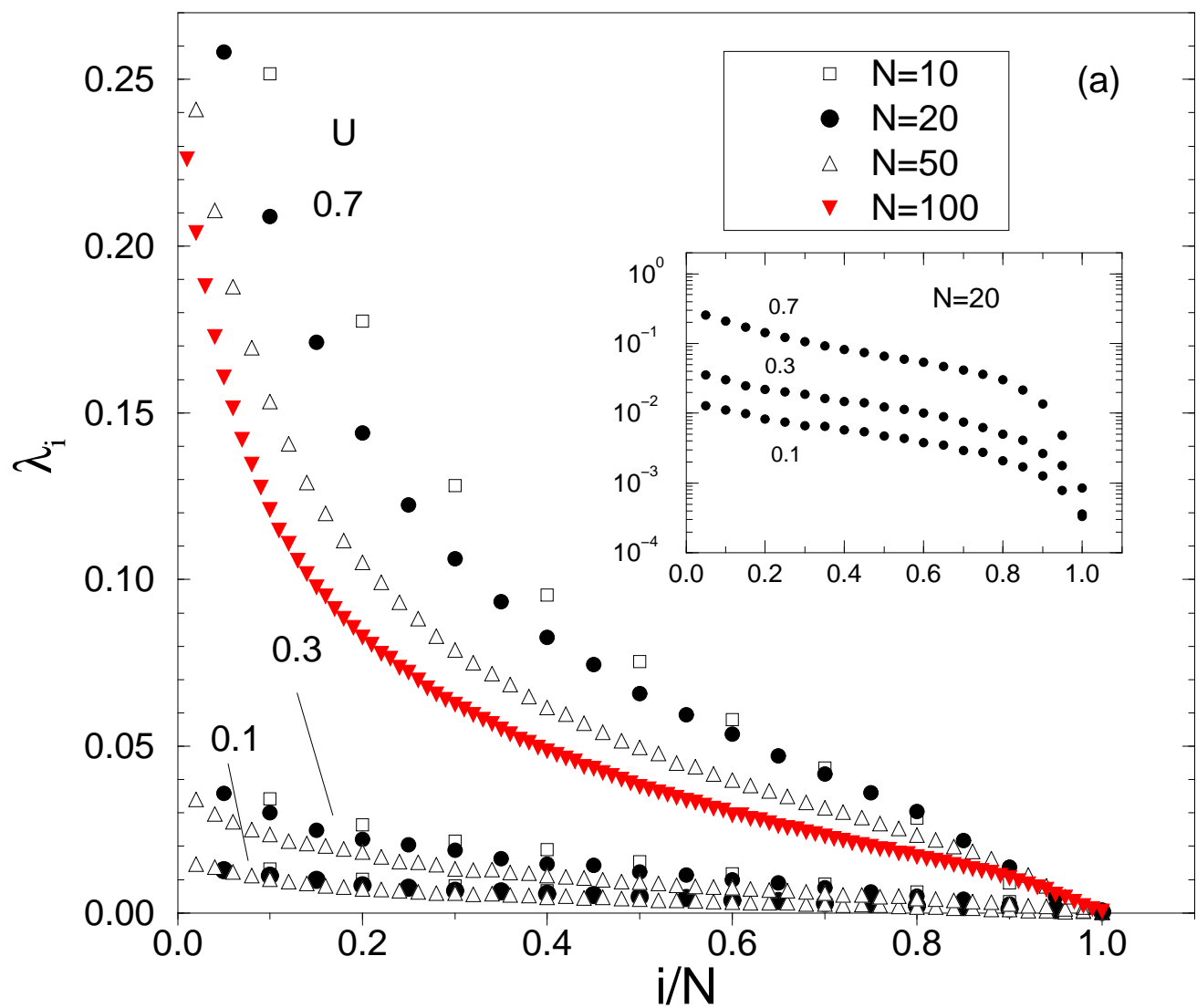
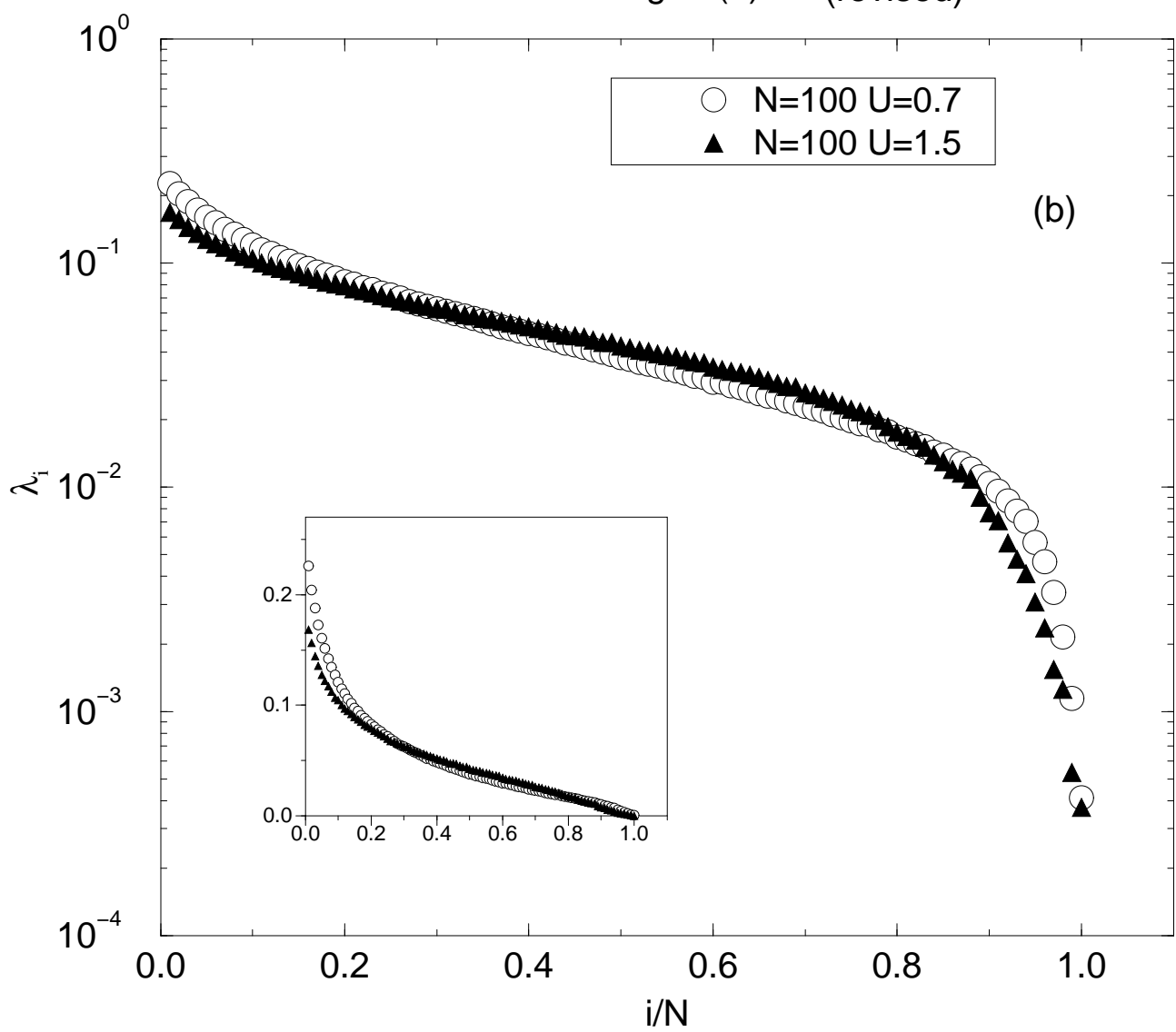
\begin{figure}
\caption{
(a)Lyapunov spectra for three different values of $U$ at 
different $N$ in linear and lin-log scale (inset for $N=20$) showing
the convergence (weaker for the $U=0.7$ case) to the asymptotic 
distribution.
(b)Lyapunov spectra for fixed $N=100$ with two different values of the energy, 
below and above the critical energy. No significative differences are
observed at these values of $N$.} 
\end{figure}

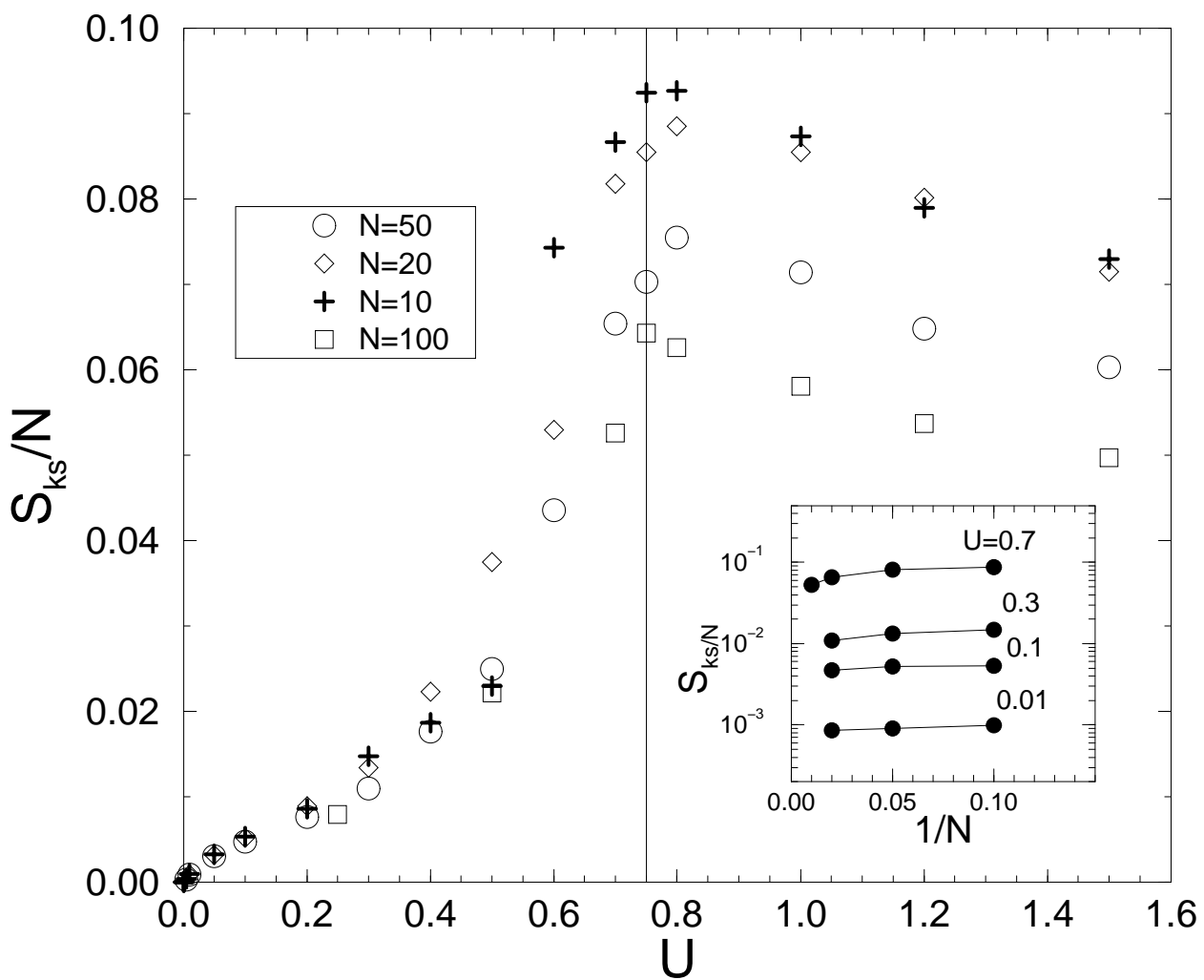
\begin{figure} 
\caption{The Kolmogorov-Sinai entropy density $S_{KS}/N$ vs. $U$ 
at increasing values of $N$. The convergence to the thermodynamic
limit of the K-S entropy density is shown in the inset for
$U=0.01$,$0.1$,$0.3$,$0.7$. The convergence is weaker for this latter
$U$ value.}
\end{figure}

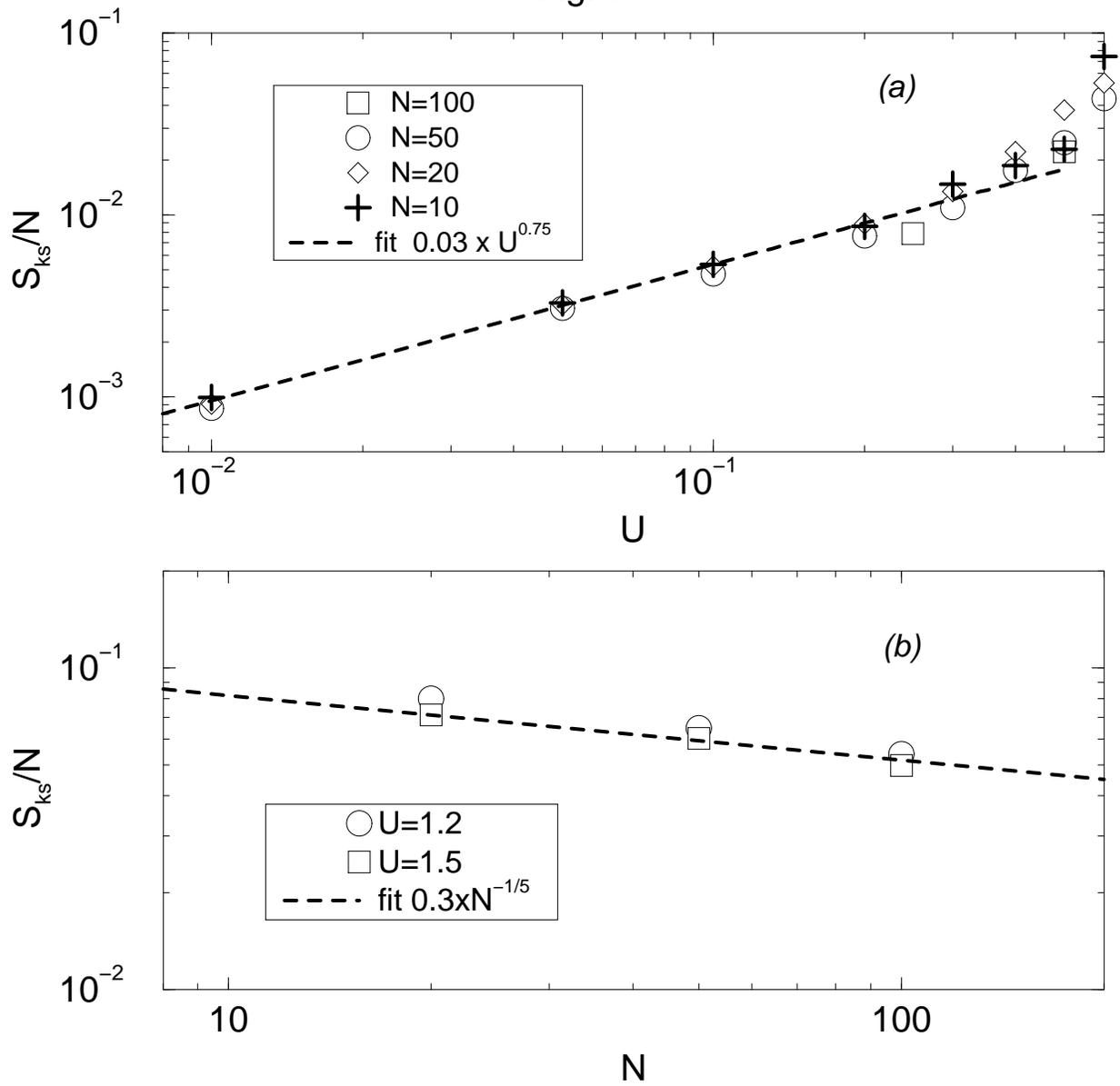
\begin{figure}
\caption{
(a) Growth of the K-S entropy density at small $U$ with the
exponent $3/4$. A rather weak $N$-dependence is observed.
(b) Convergence to zero of the K-S entropy density as $N$ 
increases in the high energy region. The fitted curve is 
the power law $N^{-1/5}$.}
\end{figure}

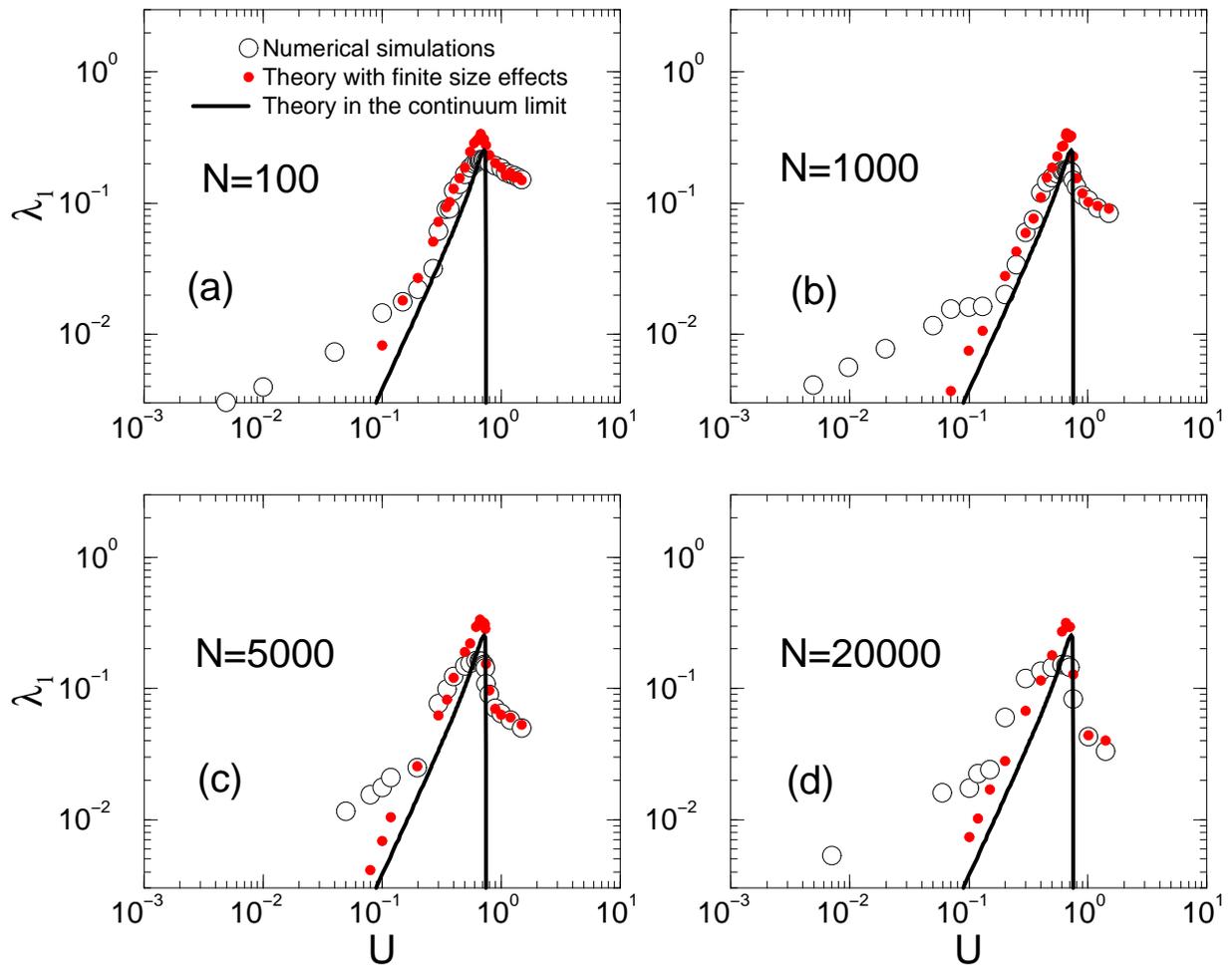
\begin{figure}
\caption{Comparison of formula (\ref{casettiformula}) for $\lambda_1$ 
at the thermodynamic limit (full line) and for finite $N$ 
(shaded circles) with numerical simulations (open circles) at increasing
values of $N$.}
\end{figure}
\vfill

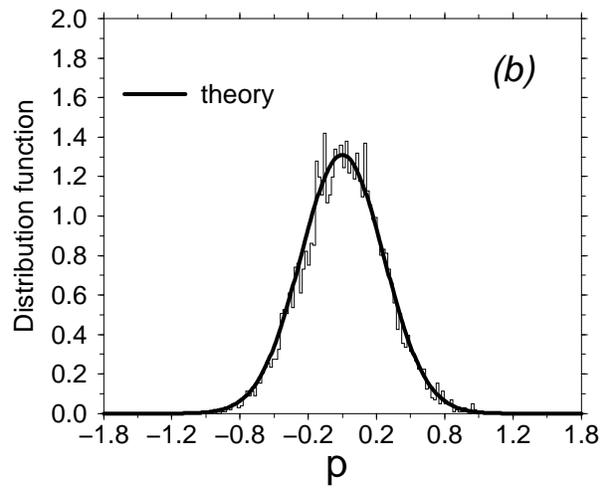
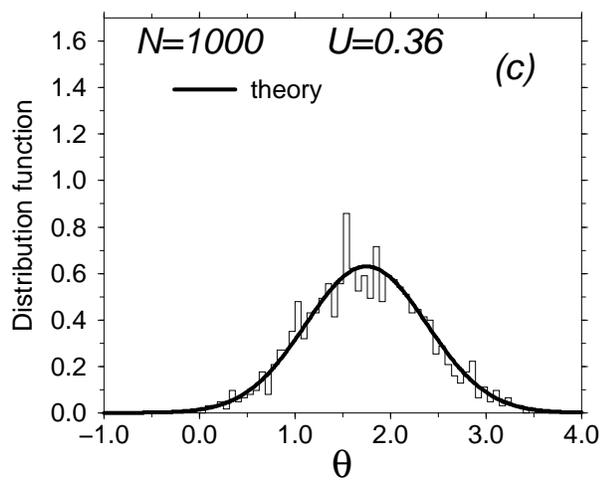
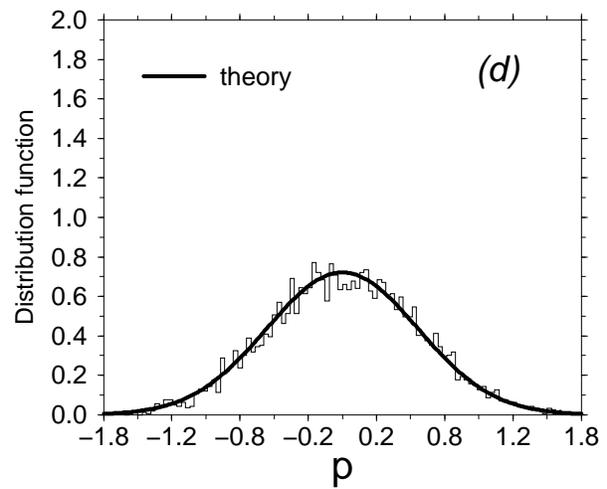
\end{document}